\documentclass[showpacs,twocolumn,amsmath,amssymb,prb,superscriptaddress]{revtex4}

\usepackage{graphicx}
\usepackage{amsmath,amsfonts,amssymb,amsthm,bm}
\usepackage{fancyhdr}
\usepackage{mathrsfs}
\usepackage{epstopdf}
\usepackage{color}
\usepackage{hyperref}

\usepackage{tikz}
\usetikzlibrary{arrows,decorations.pathmorphing,decorations.markings,decorations.shapes,backgrounds,positioning,fit,trees,calc} 
\tikzset{snake it/.style={decorate, decoration=snake}}
\usepackage{feynmp}

\tikzset{->-/.style={decoration={
  markings,
  mark=at position #1 with {\arrow{>}}},postaction={decorate}}}
\tikzset{
   vertex/.style={circle, inner sep=0pt, minimum size=5pt,fill=black,label=#1}, vertex/.default=\text{},
   crossing/.style={circle, inner sep=0, minimum size=0,label=#1}, crossing/.default=\text{},
   named/.style={draw,circle, inner sep=0pt, minimum size=12pt},
   namedE/.style={draw,circle, inner sep=0pt, minimum size=7pt},
   bath/.style={draw,thick,->-=.5}, bath/.default={},
   bathr/.style={draw,thick,dashed,->-=.5},
   time/.style={draw,dashed,thin},
   system/.style={draw,thick}, system/.default={},
   syslabel/.style={midway,auto,green!40!black}}
\tikzset{dotted pattern/.style args={#1 and #2}{
   decorate,
   fill=black,
   decoration={
    shape backgrounds,
    shape=circle,
    shape size=#1,
    shape sep={#2, between center},
    }
  },
  dotted pattern/.default={1pt and 1.5mm},
}

\begin{document}


\title{Diagrammatic description of a system coupled strongly to a bosonic bath}

\author{Michael Marthaler}
\affiliation{Institut f\"ur Theoretische Festk\"orperphysik, Karlsruhe Institute of Technology, D-76128 Karlsruhe, Germany}

\author{Juha Lepp\"akangas}
\affiliation{Institut f\"ur Theoretische Festk\"orperphysik, Karlsruhe Institute of Technology, D-76128 Karlsruhe, Germany}




\date{\today}

\begin{abstract}
We study a system-bath description in the strong coupling regime
where it is not possible to derive a master equation for the reduced density matrix by a direct expansion in the system-bath coupling.
A particular example is a bath with significant spectral weight at low frequencies.
Through a unitary transformation it can be possible to find a more suitable small expansion parameter.
Within such approach we construct a formally exact expansion of the master equation on the Keldysh contour. We consider a system diagonally coupled to a bosonic bath and expansion in terms of a non-diagonal hopping term.
The lowest-order expansion is  equivalent to the so-called $P(E)$-theory or non-interacting blip approximation (NIBA).
The analysis of the higher-order contributions shows that there are two different classes of higher-order diagrams.
We study how the convergence of this expansion depends on the form of the spectral function with significant weight at zero frequency.
\end{abstract}

\maketitle


\section{Introduction}
System-bath approaches are commonly used in many fields of physics~\cite{Buch_Weiss}.
Particularly important they are for quantum optics and quantum transport~\cite{Buch_Carmichael}.
Generally, in a system-bath approach one defines  a 'system' which contains a small number of degrees of freedom and a 
'bath' which contains a large number of degrees of freedom.
The system and the bath are mostly coupled linearly, although also different forms of coupling are possible.

A common approach to solve this problem is to expand the time evolution of the full
density matrix in the coupling between the system and the bath and trace out the bath degrees of freedom.
This results in an effective equation of motion for the (reduced) density matrix of the system. 
The well known examples are the Bloch-Redfield~\cite{Shnirman_Review} and Lindblad master equations~\cite{Lindblad}.
Both of these models require a weak coupling between the system and the bath. To be more precise, they 
require a bath correlator with a decay rate which is large compared to the effective system-bath coupling. 
However, if the system-bath coupling is strong, other approaches can be more useful.
For two-level systems a well known approach is the
polaron transformation~\cite{Polaron_in_second_order,Polaron_strong_coupling,Polaron_subohmic},
i.e.~transformation into the bath-dressed system-states.
This has first been introduced to study polaronic hopping of electons with wavefunctions confined to single atomic sites~\cite{Polaron_Review}.
Using the polaron transformation, the bath contribution is completely diagonalised. It is then
necessary to expand in the terms of the original system Hamiltonian that do not commute with the system-bath
coupling operator, dressed now by the bath operators.
This approach is also known as non-interacting blip approximation~\cite{NIBA_Original,NIBA_Thorwart,NIBA_Wilhlem} (NIBA)
and can be extended to the weakly interacting blip approximation~\cite{WIBA_Thorwart_v1,WIBA_Thorwart_v2}.

It is not always straightforward to define the coupling strength between the system and the bath.
In many cases it is helpful to consider the spectral function of the bath modes to understand which method might work best. 
If the spectral function is very smooth, a direct expansion in the system-bath coupling is usually warranted. However, 
if the spectral function has sharp peaks more care is necessary. A simple example is the spectral function
$S(\omega)=\gamma\lambda^2/[\lambda^2+(\omega-\omega_0)^2]$ (we use the notation $\hbar=1$).
If an energy splitting of the system is close to the peak at $\omega=\omega_0$,
it is possible to consider the height of the peak ($\gamma$) as the strength of the coupling, whereas the width
of the peak ($\lambda$) gives us a good indication of the decay time of the bath correlator.
Then, for $\gamma/\lambda\gg 1$ we are in the strong-coupling limit and for $\gamma/\lambda\ll 1$ we are in the weak-coupling limit.

In this article, we investigate a situation with a bath spectral density which has a substantial spectral weight at low frequencies. 
If the effect of temperature is considered, we will have a spectral function  peaked at zero frequency which is exactly in the strong coupling limit,
i.e.~its peak is higher than it is wide. 
As we will discuss later, this spectral function can be a result of coupling a quantum system to a large Ohmic resistor~\cite{pofe_Nazarov}
in which case we have an Ohmic spectral density with small cut-off frequency.
We will also discuss how our noise spectrum can be considered as a subcomponent of $1/f$-noise~\cite{Shnirman_one_over_f}. 
Furthermore, the spectral density relevant for our work has been measured, e.g. in flux qubits which are used in the D-Wave devices~\cite{dwave_noise_measurment,dwave_high_frequency}. 
In larger coupled systems containing these qubits, a description based on polaronic hopping has also already been studied~\cite{Dykman_dwave}.

This article represents a continuation of our work on lasing in systems under strong noise~\cite{JC_1,JC_3} and incoherent Cooper-pair tunneling in Josephson junction arrays~\cite{JaredJJarray}. In both cases we used the lowest-order results of our expansion theory, and here we present for the first time the higher-order expansion which as used to analyze the convergence conditions. 
The same physics also govern inelastic Cooper-pair tunneling across voltage-biased Josephson junctions
in the Coulomb blockade regime~\cite{pofe_Nazarov}. In this system, higher-order diagrams similar to what considered in this article have been 
earlier formulated by us and also explicitly evaluated beyond the leading order~\cite{Leppakangas2015,Leppakangas2016}.

This paper is organized as follows.
In Sec.~\ref{sec:Model}, we start by introducing our model in an abstract way and in Sec.~\ref{sec:PhysicalRealisations} we discuss specific physical realizations where this expansion is applicable and has partially already been used.
In Sec.~\ref{sec:KeldyshContour}, we introduce our expansion theory to all orders on the Keldysh contour. In Sec.~\ref{sec:TheNoiseSpectra},   we study the convergence conditions in the specific case if strong low-frequency noise, which can of course also be used to estime the validity of the lowest order expansion. We consider both the low- and high-temperature regimes.
The conclusions and discussion are given in Sec.~\ref{sec:Conclusions}.


\section{The Model}\label{sec:Model}

We consider a
system coupled diagonally to a bosonic bath in the presence of a non-diagonal hopping term,
described by the total Hamiltonian
\begin{equation}\label{eq_original_Hamiltonian}
 H_0=H_D+(\hat{T}+\hat{T}^{\dag})+\hat{D}\sum_i c_i (\hat b_i^{\dag}+\hat b_i) + \sum_i \omega_i \hat b_i^{\dag}\hat b_i.
\end{equation}
The system Hamiltonian consists of $H_D$, a Hamiltonian which commutes with the coupling operator $\hat{D}$,
and a (hopping) part which does not commute with $\hat{D}$, given by $\hat{T}+\hat{T}^{\dag}$.
The system is coupled linearly to the bath of bosonic modes with frequencies $\omega_i$,
described by
the corresponding annihilation (and creation) operators $\hat b_i^{(\dag)}$.
Here, the commutator between the operators $\hat{T}$ and $\hat{D}$ satisfies the property
\begin{equation}\label{eq:TheoryCondition}
 [\hat{T},\hat{D}]=c \hat{T},
\end{equation}
where $c$ is a constant. In Sec.~\ref{sec:PhysicalRealisations}, we introduce several systems whose Hamiltonians satisfy the abovementioned properties.

Provided by Eq.~(\ref{eq:TheoryCondition}), it is now convenient to do the unitary (polaron) transformation
\begin{equation}
 U_D=\exp\left[-\hat{D}\sum_i \frac{c_i}{\omega_i}(\hat b_i^{\dag}- \hat b_i)\right],
\end{equation}
and bring the resulting total Hamiltonian into the standard form of a system-bath approach,
\begin{equation}\label{eq:Hamiltonian1}
 H=U_D^{\dag}H_0 U_D=H_{\rm S}+H_{\rm C}+H_{\rm B},
\end{equation}
with
\begin{eqnarray}
H_{\rm S} &=& H_D-\hat{D}^2\sum_i\frac{c_i^2}{\omega_i}\\ 
H_{\rm C} &=& \hat{T} e^{-\hat \phi}+\hat T^{\dag} e^{\hat\phi}\,\, ,\,\, \phi=c\sum\frac{c_i}{\omega_i}(\hat b_i^{\dag}-\hat b_i)\label{eq:HamiltonianC}  \\
H_{\rm B} &=& \sum_i \omega_i \hat b_i^{\dag} \hat b_i.\label{eq:Hamiltonian4}
\end{eqnarray}
The system Hamiltonion $H_{\rm S}$ consists of $H_D$ (introduced above) and a renormalisation coming from the bath.
Coupling to the bath is described by $H_{\rm C}$ and the bath Hamiltonian $H_{\rm B}$ remains unchanged.
It will now be our goal to derive a master equation by expanding the equation of motion for the reduced density matrix in orders of $H_C$.
For example, when applied to the spin-boson model
or to the Josephson junction coupled to an electromagnetic environment,
this corresponds to an expansion of the system dynamics in powers of the tunneling coupling, as discussed below.


\section{Physical realisations}\label{sec:PhysicalRealisations}
In this section, we introduce specific physical realizations where our expansion is applicable.
The specific form of diagrammatic expansion, the leading-order master equation, and its convergence analysis are then given in Secs.~\ref{sec:KeldyshContour} and~\ref{sec:TheNoiseSpectra}

Generally,
prior to the unitary transformation, the system Hamiltonian and the coupling Hamiltonian can be described by diagonal contributions 
\begin{equation}
H_D=\sum_n \epsilon_n |n\rangle\langle n| ,
\end{equation}
and similarly for the coupling to the bath
\begin{equation}
\hat D=\sum_n d_n |n\rangle \langle n|.
\end{equation}
In the following,
we will consider an off-diagonal part with (hopping) coupling between the nearest system levels
\begin{equation}
\hat T= \sum_n {\tau}_n |n\rangle \langle n+1| .
\end{equation}
For our theory to be applicable, that is Eq.~(\ref{eq:TheoryCondition}) to be valid,
the matrix elements $d_n$ need to have the property $d_{n+1}-d_n=c$.
We point out that the approach can also be extended to more complicated operators $\hat T$.
Let us now discuss some important models from the literature which can be mapped to the Hamiltonian of Eq.~(\ref{eq_original_Hamiltonian}).

\subsection{Spin-Boson Model}\label{subsec_spinboson}

One of the most well studied problems in system-bath physics is the spin-boson model~\cite{Spin_Boson_1,Spin_Boson_2}.
It describes many interesting systems and phenomena, like electron-transfer reactions~\cite{Spin_Boson_3},
bio molecules~\cite{Spin_Boson_4}, cavity QED~\cite{Spin_Boson_5,Spin_Boson_6} and general dissipative quantum systems \cite{Spin_Boson_7,Spin_Boson_8}.
Here we have the Hamiltonian
\begin{eqnarray}
 H&=&\frac{1}{2}\epsilon \hat\sigma_z - \frac{1}{2}\Delta\hat\sigma_x \nonumber \\
 & &+ \hat\sigma_z \sum_i c_i (\hat b_i^{\dag}+\hat b_i) + \sum_i \omega_i \hat b_i^{\dag}\hat b_i.
\end{eqnarray}
Here $\hat \sigma_i$ are the Pauli-matrices acting on a two-level system.
This can be mapped onto the Hamiltonian of Eq.~(\ref{eq_original_Hamiltonian}),
with identification
\begin{equation}
H_D= \frac{1}{2}\epsilon\hat\sigma_z \,\,\,\,\, , \,\,\,\,\, \hat{T}=-\frac{1}{2}\Delta\hat\sigma_+ \,\,\,\,\, , \,\,\,\,\, D=\hat\sigma_z.
\end{equation}
Here we introduced the spin raising and lowering operators $\hat\sigma_+ + \hat\sigma_- = \hat\sigma_x$.
For the spin-boson model an expansion in terms of $\hat{T}$ in lowest order is known as the non-interacting blip approximation~\cite{NIBA_Original}.
Also higher order expansions have been formulated~\cite{WIBA_Thorwart_v1,WIBA_Thorwart_v2}.

\subsection{ Jaynes-Cummings Model }

The Jaynes-Cummings model describes an interaction between a single electromagnetic mode and a two-level system. 
If the two-level system is coupled to a bosonic bath (eg.~to describe decoherence) the total Hamiltonian is given by
\begin{eqnarray}
 H &=& \frac{1}{2}\epsilon \hat\sigma_z+g(\hat \sigma_+ \hat a+\hat\sigma_- \hat a^{\dag})+\omega \hat a^{\dag}\hat a  \nonumber \\
  & & +\hat \sigma_z \sum_i c_i (\hat b_i^{\dag}+\hat b_i) + \sum_i \omega_i \hat b_i^{\dag} \hat b_i.
\end{eqnarray}
 This model has been studied by us in the context of inversionless lasing \cite{JC_1}
 and coupling of quantum dots to a transmission-line resonator \cite{JC_2,JC_3}.
 We have here
\begin{equation}
H_D=\frac{1}{2}\epsilon\hat\sigma_z+\omega \hat a^{\dag} \hat a \,\,\,\, , \,\,\,\,  \hat T=g \hat a\hat \sigma_+ \,\,\,\, , \,\,\,\,  \hat D=\hat \sigma_z .
\end{equation} 
 When expanding in $\hat{T}$
 one has to note that the expansion parameter grows with the photon number.
 Therefore it is clear that a lowest-order approximation is only valid for small photon numbers
 (small $g\sqrt{n}$ compared to $\omega$, where $n$ is the resonator photon number).

\subsection{Superconducting devices in the charge regime}
Another systems of great interest where the above discussion is valid are superconducting devices in the charge regime.
As an example, the Hamiltonian of a superconducting charge qubit connected
(capacitively) to a transmission line can be written in the form,
\begin{eqnarray}
 H &=& E_C \hat{N}^2 -E_{\rm J}\cos\hat{ \theta}  \\
   &+& \hat{N}   \sum_i c_i (\hat b_i^{\dag}+\hat b_i) + \sum_i \omega_i \hat b_i^{\dag}\hat b_i .\nonumber
\end{eqnarray}
Here $\hat N$ is the excess Cooper-pair number on the island with Cooper-pair charging energy $E_C$, and
$E_{\rm J}$ is the Josephson coupling
describing Cooper-pair tunneling between the island and the lead.
The superconducting phase and the charge operator are conjugate variables
and satisfy $[\hat{N},e^{i\hat{\theta}}]= e^{i\hat{\theta}}$, which means that we can identify,
\begin{equation}
H_D= E_C \hat{N}^2  \,\,\, , \,\,\,  \hat{T}=-E_{\rm J} e^{i\hat{\theta}}/2 \,\,\, , \,\,\,  \hat{D}=\hat{N}.
\end{equation}
For this case the lowest-order expansion is equivalent to the $P(E)$-theory~\cite{pofe_Nazarov}.
In this system also expansion schemes to higher-orders in $E_{\rm J}$ have been considered
by us and others~\cite{Ingold1998,Ingold1999,Leppakangas2015,Leppakangas2016}.
Naturally many other noise sources, like sub-gap quasiparticles \cite{QuasiJuha,QuasiHeimes},
can have an effect on supercondcuting systems. 
In the limit of large $E_{\rm J}$ the noise characteristics of this model can also change substantially \cite{DimaNoise} and for highly 
structured environments open system methods have been discussed~\cite{Juha_Open}.

\subsection{Multi-partite systems}

The model presented here can also easily be extended to include coupling of many system operators
to independent baths. Here each of the system operators has to have a similar relation with the system Hamiltonian as 
 in Eq.~(\ref{eq:TheoryCondition}). In particular, this model can be used to study incoherent Cooper-pair tunneling in 
Josephson junction arrays~\cite{ZimmerJJarray,JaredJJarray}
and it could also be useful when considering hopping between many
coupled two-level systems with low coherence~\cite{DWAVE_Troyer,Pascal_Meta}.

\section{Expansion on the Keldysh contour}\label{sec:KeldyshContour}
 
The total Hamiltonian $H$ is divided into three parts:  the quantum system $H_{\rm S}$,
 the bath $H_{\rm B}$, and the coupling between the quantum system and the bath $H_{\rm C}$, see Eqs.~(\ref{eq:Hamiltonian1}-\ref{eq:Hamiltonian4}).
Our aim now is to derive an equation of motion for the reduced density matrix for the quantum system,
where we trace out the degrees of freedom of the bath.
The expansion of the time evolution on Keldysh contour
is dicussed extensively in the literature, see for example Ref.~[\onlinecite{SchoellerSchoenKeldysh}]. Below,
we will give a short review of the relevant steps. 
Differences to usual approaches appear
when we introduce the contraction method of exponentialized bosonic operators 
(Sec.~\ref{sec:ContractionRules}).

\subsection{Time evolution of the reduced density matrix} 
We start with the equation of motion for the average value of the projection operator $\hat{P}_{ss'}$,
where $|s\rangle$ are the eigenstates of $H_{\rm S}$. We have then
\begin{equation}
\hat{P}_{ss'}=|s'\rangle \langle s| \,\,\,\, , \,\,\,\,  H_{\rm S}|s\rangle=E_s|s\rangle.
\end{equation}
In this notation we can define the elements of the reduced density matrix as
\begin{equation}
P_{ss'}(t)=\left\langle \hat{P}_{ss'}(t)\right\rangle.
\end{equation}
The time evolution is then given by,
\begin{equation}\label{eq_Poft_before_expansion}
P_{ss'}(t) = {\rm Tr}\left[\hat \rho(t_0)\hat U_{\rm I}^\dagger(t,t_0) \hat{P}_{ss',{\rm I}}(t) \hat U_{\rm I}(t,t_0)\right].
\end{equation}
This approach is equivalent to the Nakajima-Zwanziger projection formula~\cite{Buch_Weiss}.
Here we use the definition of an operator $\hat{O}$ in the interaction picture
\begin{equation}
\hat O_{\rm I}(t)=e^{i(H_{\rm S}+H_{\rm B})(t-t_0)}\hat{O} e^{-i(H_{\rm S}+H_{\rm B})(t-t_0)}.
\end{equation}
The time-evolution operator  in the interaction picture is given by
 \begin{equation}\label{eq_time_evolution_operator_in_the_interaction_picture}
    U_{\rm I}(t,t_0)={\cal{T}} e^{-i\int_{t_0}^t H_{ \rm C,I}(t') dt'}
 \end{equation}
where $\cal{T}$ is the time-ordering operator ($t>t_0$).

We assume now that at time $t_0$ the density matrix separates into the density matrix of the bath $\hat\rho_{\rm B}(t)$ and system $\hat\rho_{\rm S}(t_0)$,
and write it in the form
\begin{equation}\label{eq_Def_of_rho_at_t0}
      \hat\rho(t_0)=\hat\rho_{\rm B}(t_0) \hat\rho_{\rm S}(t_0)=\hat\rho_{\rm B} \sum_{\bar{s}\bar{s}'} P_{\bar{s}\bar{s}'}(t_0) |\bar{s}\rangle \langle \bar{s}'| .
\end{equation}
Combining Eqs.~(\ref{eq_Poft_before_expansion}) and (\ref{eq_Def_of_rho_at_t0})  allows us to write,
     \begin{equation}
      P_{ss'}(t)=\sum_{\bar{s}\bar{s}'} P_{\bar{s}\bar{s}'}(t_0) \Pi_{\bar{s}\bar s'\rightarrow ss'}(t_0,t),
     \end{equation}
with the time-evolution of the super-operator $\Pi(t_0,t)$.
Expanding the time-evolution operators as in Eq.~(\ref{eq_time_evolution_operator_in_the_interaction_picture}) gives us
       %
   \begin{eqnarray}\label{eq_Pi_expandedgeneral}
   & &\Pi_{\bar{s}\bar s'\rightarrow ss'}(t_0,t)  \\
   & & = \langle \bar{s}'|{\rm Tr}_{\rm B} \rho_{\rm B}(t_0)
               \sum_{m=0}^{\infty}
               (-i)^m \int_{t_0}^{t} dt_1'\int_{t_0}^{t_1'} dt_2'\ldots\nonumber \\  
            &  &\!\int_{t_0}^{t_{m-1}'}\!\!  dt_m'\! 
              {\cal{T}}_K\!\!\left( \! H_{\rm C,I}(t_1')\! H_{\rm C,I}(t_2')\! \right.\!\ldots
             \left. \! H_{\rm C,I}(t_m')\!\hat{P}_{ss',{\rm I}}(t)\!\right)\!
              |\bar{s}\rangle\nonumber \, .
    \end{eqnarray}
 Here $\cal{T}_K$ represents the time sorting {\em along the Keldysh contour},
 which we will explain below.

 The Keldysh contour has two branches. The upper branch represents the time evolution from $t_0$ to $t$, and the lower branch represents time-evolution in the opposite direction. In our case the time 
 $t$ is determined by the projection operator $\hat{P}_{ss',{\rm I}}(t)$. All operators to the right of $\hat{P}_{ss',{\rm I}}(t)$ will be on the upper branch of the Keldysh contour,
 all operators to the left will be on the lower branch. Each lower-branch operator
 will be associated with an extra factor $-1$.

 In the coupling Hamiltonian $H_{\rm C}$ we have the operators $\hat T e^{-\hat\phi(t)} $ and  $\hat T^{\dag} e^{\hat\phi(t)}$, see Eq.~(\ref{eq:HamiltonianC}). 
 We separate these two terms in our description of the expansion on Keldysh contour.
 The operators $\hat T e^{-\hat\phi(t)} $ are represented by filled circles, while $\hat T^{\dag} e^{\hat\phi(t)}$ by empty circles.  
 On Keldysh contour the time-evolution of the super-operator is then given by 
 \begin{eqnarray}\label{diagram:First}
 \Pi(t_0,t)=
          \sum_{m=0}^{\infty} i^m
    \begin{tikzpicture}[anchor=base,baseline=8pt]
    \coordinate (A) at (0,0);
    \coordinate (B) at (2.8,0);
    \coordinate (C) at (0,0.7);
    \coordinate (D) at (2.8,0.7);
    \coordinate (D1) at (0.5,0);
    \node (N1) at (0.5,-0.3) {$t_m$};
    \coordinate (D2) at (1.2,0);
    \node (N2) at (1.2,-0.3) {$t_{m-2}$};
    \coordinate (D3) at (2.2,0);
    \node (N3) at (2.2,-0.3) {$t_{2}$};
    \coordinate (D4) at (0.85,0.7);
    \node (N4) at (0.85,0.9) {$t_{m-1}$};
    \coordinate (D5) at (1.9,0.7);
    \node (N5) at (1.9,0.9) {$t_{3}$};
    \coordinate (D6) at (2.4,0.7);
    \node (N6) at (2.4,0.9) {$t_{1}$};
    \node (LD) at (1.6,0.28) {$\ldots$};
    \node (E) at (3.15,0.28) {$\times$};
    \node (T) at (3.15,0.7) {$t$};
    \node (T0) at (0,0.9) {$t_0$};
    \draw[line width=1.0pt] (A) -- (B);
    \draw[ line width=1.0pt,dashed] (A) -- (C);
     \path[system] (B.east) arc[start angle=-90, delta angle=90, y radius = 0.35cm, x radius = 0.35cm] 
  node[above right] {$s$} node[below right] {$s'$} arc[start angle=0, delta angle=90, y radius = 0.35cm, x radius = 0.35cm];
    \draw[line width=1.0pt] (D) -- (C);
    \fill (D1) circle (2pt);
    \fill[white] (D1) circle (1pt);
    \fill (D2) circle (2pt);
    \fill (D3) circle (2pt);
    \fill (D4) circle (2pt);
    \fill (D5) circle (2pt);
    \fill[white] (D5) circle (1pt);
    \fill (D6) circle (2pt);
    \fill[white] (D6) circle (1pt);
    \end{tikzpicture}
 \end{eqnarray}
 Here, for each order $m$ a summation over all geometrically different diagrams is made.

To define a self-energy we will now establish contraction rules for correlators
 of the form
\begin{eqnarray}
  & &{\rm Tr}_{\rm B}  \left[\hat \rho_{\rm B} (t_0) {\cal{T}}_K e^{\hat \phi_1} e^{-\hat \phi_2} e^{\hat \phi_3} \ldots e^{\hat \phi_m}\right] \\
  & & = \left\langle {\cal{T}}_K e^{\hat \phi_1} e^{-\hat \phi_2} e^{\hat \phi_3} \ldots e^{\hat \phi_m}\right\rangle_{\rm B} \nonumber \\
 &&= \left\langle e^{\hat \phi_m}e^{-\hat \phi_{m-2}} \ldots e^{-\hat \phi_2} e^{\hat \phi_1} e^{\hat \phi_3} \ldots e^{-\hat \phi_{m-1}}\right\rangle_{\rm B}, \nonumber
\end{eqnarray}
where the time-ordering of the lowest line corresponds to the diagram in Eq.~(\ref{diagram:First}).
 We have used the notation $\hat \phi(t_i)\equiv \hat \phi_i$. It is the next step which is different
 from usual master equation derivations.
 This is since the  Wick's theorem does not apply to operators of the type $\exp\left[{\hat \phi_i}\right]$
 and we cannot rewrite this $m$-time correlator in products of two-time correlators in the usual way.

\subsection{Contraction rules}\label{sec:ContractionRules}

The Feynman disentangling method allows us to derive a helpful simplification
for ensemble averages of products of operators $\exp\left( \hat\phi_n \right)$, where $\hat\phi_n$ is an arbitrary linear combination of bosonic
annihilation and creation operators. The disentangling has the form
\begin{eqnarray}\label{eq_basicdisentrule}
    &&\left\langle e^{n_1\hat\phi_1}e^{n_2\hat\phi_2}\ldots e^{n_m\hat\phi_m}\right\rangle\\
    &&=e^{\frac{1}{2}{\langle\left(\sum_{i=1}^{m}n_i\hat\phi_i\right)^2 \rangle}}
      e^{\frac{1}{2}\sum_{i<j}^{m}\left[n_i\hat\phi_i,n_j\hat\phi_j\right]}\, . \nonumber
 \end{eqnarray}
    %
    Here, the factors $n_i$ take values $\pm 1$.
	This result can also be derived by applying the Wick's theorem to Taylor expansions of the exponentialized operators, similarly as in Ref.~[\onlinecite{pofe_Nazarov}].
    Using Eq.~(\ref{eq_basicdisentrule})
    it is straightforward to show that
    averaging products over the reservoir is only non-zero if there is the
    same number of operators with opposite signs of the exponents. 
    Therefore, we only have to consider
    diagrams with equal amount of filled and empty circles.
    For such combinations we can write
   \begin{eqnarray}\label{eq:ContractionPairing}
   & &\left\langle {\cal T}_K\Pi_n e^{\hat\phi_n}e^{-\hat\phi_{n'}}\right\rangle \nonumber \\
    &=& e^{\frac{1}{2} \left\langle[ \sum_n(\hat\phi_n-\hat\phi_{n'})]^2  \right\rangle}
   e^{-\frac{1}{2}\sum_{n<n'}{\cal T}_K[\hat\phi_n,\hat\phi_{n'}]} \\
    &\times& e^{\frac{1}{2}\sum_{n<m}{\cal T}_K[\hat\phi_n,\hat\phi_m]}e^{\frac{1}{2}\sum_{n'<m'}{\cal T}_K[\hat\phi_{n'},\hat\phi_{m'}]}\, .\nonumber
    \end{eqnarray}
	Here we use the notation where the number $i$ ($i'$) corresponds to a positive
	(negative) signed phase operator.

   In the next step we group  all $e^{\hat\phi_n}$ to the $e^{-\hat\phi_{n'}}$ closest to each other on the real time axis.
   The difference to usual diagramatic formulations is that each circle is paired only once.
   This is done practically by grouping the timewise earliest empty circle to timewise earliest filled circle, and so on.
   This is the only possible way to connect the diagrams that allows for a consistent definition of a selfenergy.
   For example, the lowest order contribution to the selfenergy will contain exactly two vertices, one empty and one filled.
   This contribution has to be repeated $n$-times for a digrammatic part which contains $n$ selfenergies.
   Therefore, we have to connect the  corresponding circles  in diagrams with  $2n$ vertices to reproduce the $n$ lowest order contractions.
   We will discuss this further below for a specific example.

   Interactions beyond the pairings are included by pair connectors, introduced below.
   To write the resulting correlators in a compact form we introduce the notation
   \begin{equation}\label{eq:FunctionTimeOrdering}
   f({\cal{T}_K}[t_1,t_2,t_3,\ldots]) =  f(t_1,t_2,t_3,\ldots) \, .
   \end{equation}
   The expression ${\cal T_K}[t_1,t_2,t_3,\ldots]$ implies that the arguments of the function $f$ should be time sorted
   along the Keldysh contour. For the particular example in Eq.~(\ref{eq:FunctionTimeOrdering}) we assumed $t_1>t_2>t_3, \ldots$ along the Keldysh contour.

   We then separate contributions from the timewise nearby pairs (introduced above)
   from other terms, that will describe interaction between these pairs.
   This allows us to bring the correlator into the form,
   \begin{eqnarray}
    & &\left\langle {\cal T}_K \Pi_n e^{\hat\phi_n}e^{-\hat\phi_{n'}}\right\rangle \label{eq:Correlator1} \\
    &=& \Pi_n\left\langle {\cal T}_K e^{\hat\phi_n}e^{-\hat\phi_{n'}}\right\rangle
    \,\Pi_{n<m} \left(F({\cal{T}}_K[t_n,t_{n'},t_m,t_{m'}])+1\right)\, \nonumber\\
     &=&\Pi_n P({\cal{T}_K}[t_n,t_{n'}]) \nonumber\\
     &+& \Pi_n P({\cal{T}_K}[t_n,t_{n'}])
     \rangle F({\cal{T}}_K[t_2,t_{2'},t_1,t_{1'}])+\ldots\, .\nonumber
    \end{eqnarray}
The two-time correlator $P(t_1,t_2)$ has the form
\begin{equation}\label{eq:Correlator}
P(t_1,t_2)=\left\langle e^{\hat\phi(t_1)}e^{-\hat\phi(t_2)}\right\rangle=e^{C(t_1-t_2)}.
\end{equation}
Here, the pair correlator $C(t)$ is related to the bath structure as
\begin{eqnarray}\label{eq:Correlator2}
C(t)&=&\frac{c^2}{\pi} \int_0^{\infty}\, d\omega
\frac{J(\omega )}{\omega^2}  \\
&\times & \left[\coth\left(\frac{\omega}{2 k_{\rm B} T}\right)(\cos\omega t  -1 ) 
- i \sin\omega t  \right]     , \nonumber
\end{eqnarray}
where we have taken the continuum limit by defining
$\pi\sum_i c_i^2 f(\omega_i)/\omega_i^2 \equiv \int d\omega J(\omega) f(\omega)/\omega^2 $,
for an arbitrary function $f(\omega)$.
The Fourier transform of Eq.~(\ref{eq:Correlator}) is known from the $P(E)$-theory~\cite{pofe_Nazarov},
\begin{equation}\label{eq:PEFunction}
P(E)=\frac{1}{2\pi}\int_{-\infty}^{\infty} dt e^{C(t) +iE} .
\end{equation} 
It satisfies $\int_{-\infty}^{\infty}P(E)dE=1$ and
describes the probability to exchange energy $E$ with the bosonic environment in
an incoherent transition between two system states connected by $\hat T$ or $\hat T^\dagger$.
As discussed more detailed in Sec.~\ref{sec:TheNoiseSpectra},
the function $J(\omega)$ is related to the real part of the
environmental impedance $Z(\omega)$ as
$ J(\omega)=2\pi\omega {\rm Re}[Z(\omega)]/R_{\rm Q}$, where $R_{\rm Q}=h/e^2$ is the
resistance quantum.

    The interaction between the pairs is described by the function 
     \begin{eqnarray}
     F({\cal T}_K [t_n,t_{n'},t_m,t_{m'}])=e^{\left\langle {\cal T}_K(\hat\phi_n-\hat\phi_{n'})(\hat\phi_m-\hat\phi_{m'})\right\rangle}-1,
     \end{eqnarray}
     where the first term on the right-hand side collects all the terms describing interaction
     between two pairs.
     Notice the addition and substraction of 1 when inserted into the full correlator in Eq.~(\ref{eq:Correlator1}):
     the contribution beyond 1 (function $F$) describes deviations from Gaussian contractions.
     This definition is sound, since at finite temperatures and in the long-time limit the function $F$ decays
     exponentially to zero with increasing time separation between the two pairs. 
 We call the function $ F({\cal T}_K[t_n,t_{n'},t_m,t_{m'}])$ {\em connector}. 


We are now ready to go forward in using our diagrammatic formulation of the problem. 
 As an example we show a contraction of an element of the time evolution, with four vertices,
  \begin{eqnarray}
    \begin{tikzpicture}[anchor=base,baseline=8pt]
    \coordinate (A) at (0,0);
    \coordinate (B) at (1.2,0);
    \coordinate (C) at (0,0.7);
    \coordinate (D) at (1.2,0.7);
    \coordinate (E) at (0.8,0.7);
    \coordinate (F) at (0.3,0.0);
    \draw[line width=1.0pt] (A) -- (B);
    \draw[line width=1.0pt] (D) -- (C);
    \fill (C) circle (2pt);
    \fill (E) circle (2pt);
    \fill (B) circle (2pt);
    \fill[white] (B) circle (1pt);
    \fill (F) circle (2pt);
    \fill[white] (F) circle (1pt);
    \end{tikzpicture}
    =\begin{tikzpicture}[anchor=base,baseline=8pt]
    \coordinate (A) at (0,0);
    \coordinate (B) at (1.2,0);
    \coordinate (C) at (0,0.7);
    \coordinate (D) at (1.2,0.7);
    \coordinate (E) at (0.8,0.7);
    \coordinate (F) at (0.3,0.0);
    \draw[line width=1.0pt] (A) -- (B);
    \draw[line width=1.0pt, snake it] (C) -- (F);
    \draw[line width=1.0pt, snake it] (E) -- (B);
    \draw[line width=1.0pt] (D) -- (C);
    \fill (C) circle (2pt);
    \fill (E) circle (2pt);
    \fill (B) circle (2pt);
    \fill[white] (B) circle (1pt);
    \fill (F) circle (2pt);
    \fill[white] (F) circle (1pt);
    \end{tikzpicture}+
    \begin{tikzpicture}[anchor=base,baseline=8pt]
    \coordinate (A) at (0,0);
    \coordinate (B) at (1.2,0);
    \coordinate (C) at (0,0.7);
    \coordinate (D) at (1.2,0.7);
    \coordinate (E) at (0.8,0.7);
    \coordinate (F) at (0.3,0.0);
    \draw[line width=1.0pt] (A) -- (B);
    \draw[line width=1.0pt, snake it] (C) -- (F);
    \draw[line width=1.0pt, snake it] (E) -- (B);
    \draw[line width=1.0pt] (D) -- (C);
    \draw[line width=1.0pt, dashed] (0.17,0.35) -- (1,0.35);
    \fill (C) circle (2pt);
    \fill (E) circle (2pt);
    \fill (B) circle (2pt);
    \fill[white] (B) circle (1pt);
    \fill (F) circle (2pt);
    \fill[white] (F) circle (1pt);
    \end{tikzpicture}
    \end{eqnarray}
    Here the dashed line describes a connector between the pair correlators (wiggled lines).
    Here we see how a diagram with four vertices is contracted in a way which reproduces the
    lowest order diagrams twice, which is necessary to allow for a consistent definition of the self energy.
    Our contraction rule is the only possible rule which allows for such a consistent definition.

 As another example we show is a contraction of an element of the time evolution with six vertices,
 \begin{eqnarray}
   \begin{tikzpicture}[anchor=base,baseline=8pt]
    \coordinate (A) at (0,0);
    \coordinate (B) at (2.,0);
    \coordinate (C) at (0,0.7);
    \coordinate (D) at (2.,0.7);
    \coordinate (D1) at (0,0);
    \coordinate (D2) at (1.,0);
    \coordinate (D3) at (1.3,0);
    \coordinate (D4) at (0.65,0.7);
    \coordinate (D5) at (1.6,0.7);
    \coordinate (D6) at (2.,0.7);
    \draw[line width=1.0pt] (A) -- (B);
    \draw[line width=1.0pt] (D) -- (C);
    \fill (D1) circle (2pt);
    \fill[white] (D1) circle (1pt);
    \fill (D2) circle (2pt);
    \fill (D3) circle (2pt);
    \fill (D4) circle (2pt);
    \fill (D5) circle (2pt);
    \fill[white] (D5) circle (1pt);
    \fill (D6) circle (2pt);
    \fill[white] (D6) circle (1pt);
    \end{tikzpicture}
    & = &
    \begin{tikzpicture}[anchor=base,baseline=8pt]
    \coordinate (A) at (0,0);
    \coordinate (B) at (2.,0);
    \coordinate (C) at (0,0.7);
    \coordinate (D) at (2.,0.7);
    \coordinate (D1) at (0,0);
    \coordinate (D2) at (1.,0);
    \coordinate (D3) at (1.3,0);
    \coordinate (D4) at (0.65,0.7);
    \coordinate (D5) at (1.6,0.7);
    \coordinate (D6) at (2.,0.7);
    \draw[line width=1.0pt] (A) -- (B);
    \draw[line width=1.0pt] (D) -- (C);
    \fill (D1) circle (2pt);
    \fill[white] (D1) circle (1pt);
    \fill (D2) circle (2pt);
    \fill (D3) circle (2pt);
    \fill (D4) circle (2pt);
    \fill (D5) circle (2pt);
    \fill[white] (D5) circle (1pt);
    \fill (D6) circle (2pt);
    \fill[white] (D6) circle (1pt);
    \draw[line width=1.0pt,snake it] (D1) -- (D4);
    \draw[line width=1.0pt,snake it] (D2) -- (D5);
    \draw[line width=1.0pt,snake it] (D3) -- (D6);
    \end{tikzpicture}
     +
      \begin{tikzpicture}[anchor=base,baseline=8pt]
    \coordinate (A) at (0,0);
    \coordinate (B) at (2.,0);
    \coordinate (C) at (0,0.7);
    \coordinate (D) at (2.,0.7);
    \coordinate (D1) at (0,0);
    \coordinate (D2) at (1.,0);
    \coordinate (D3) at (1.3,0);
    \coordinate (D4) at (0.65,0.7);
    \coordinate (D5) at (1.6,0.7);
    \coordinate (D6) at (2.,0.7);
    \draw[line width=1.0pt] (A) -- (B);
    \draw[line width=1.0pt] (D) -- (C);
    \fill (D1) circle (2pt);
    \fill[white] (D1) circle (1pt);
    \fill (D2) circle (2pt);
    \fill (D3) circle (2pt);
    \fill (D4) circle (2pt);
    \fill (D5) circle (2pt);
    \fill[white] (D5) circle (1pt);
    \fill (D6) circle (2pt);
    \fill[white] (D6) circle (1pt);
    \draw[line width=1.0pt,snake it] (D1) -- (D4);
    \draw[line width=1.0pt,snake it] (D2) -- (D5);
    \draw[line width=1.0pt,snake it] (D3) -- (D6);
    \draw[line width=1.0pt, dashed] (0.22,0.35) -- (1.25,0.35);
    \end{tikzpicture}\\
     & + & 
      \begin{tikzpicture}[anchor=base,baseline=8pt]
    \coordinate (A) at (0,0);
    \coordinate (B) at (2.,0);
    \coordinate (C) at (0,0.7);
    \coordinate (D) at (2.,0.7);
    \coordinate (D1) at (0,0);
    \coordinate (D2) at (1.,0);
    \coordinate (D3) at (1.3,0);
    \coordinate (D4) at (0.65,0.7);
    \coordinate (D5) at (1.6,0.7);
    \coordinate (D6) at (2.,0.7);
    \draw[line width=1.0pt] (A) -- (B);
    \draw[line width=1.0pt] (D) -- (C);
    \fill (D1) circle (2pt);
    \fill[white] (D1) circle (1pt);
    \fill (D2) circle (2pt);
    \fill (D3) circle (2pt);
    \fill (D4) circle (2pt);
    \fill (D5) circle (2pt);
    \fill[white] (D5) circle (1pt);
    \fill (D6) circle (2pt);
    \fill[white] (D6) circle (1pt);
    \draw[line width=1.0pt,snake it] (D1) -- (D4);
    \draw[line width=1.0pt,snake it] (D2) -- (D5);
    \draw[line width=1.0pt,snake it] (D3) -- (D6);
    \draw[line width=1.0pt, dashed] (0.25,0.25) -- (1.21,0.25);
    \draw[line width=1.0pt, dashed] (1.47,0.45) -- (1.87,0.45);
    \end{tikzpicture}
    +
      \begin{tikzpicture}[anchor=base,baseline=8pt]
    \coordinate (A) at (0,0);
    \coordinate (B) at (2.,0);
    \coordinate (C) at (0,0.7);
    \coordinate (D) at (2.,0.7);
    \coordinate (D1) at (0,0);
    \coordinate (D2) at (1.,0);
    \coordinate (D3) at (1.3,0);
    \coordinate (D4) at (0.65,0.7);
    \coordinate (D5) at (1.6,0.7);
    \coordinate (D6) at (2.,0.7);
    \draw[line width=1.0pt] (A) -- (B);
    \draw[line width=1.0pt] (D) -- (C);
    \fill (D1) circle (2pt);
    \fill[white] (D1) circle (1pt);
    \fill (D2) circle (2pt);
    \fill (D3) circle (2pt);
    \fill (D4) circle (2pt);
    \fill (D5) circle (2pt);
    \fill[white] (D5) circle (1pt);
    \fill (D6) circle (2pt);
    \fill[white] (D6) circle (1pt);
    \draw[line width=1.0pt,snake it] (D1) -- (D4);
    \draw[line width=1.0pt,snake it] (D2) -- (D5);
    \draw[line width=1.0pt,snake it] (D3) -- (D6);
    \draw[line width=1.0pt, dashed] (0.25,0.2) -- (1.21,0.2);
    \draw[line width=1.0pt, dashed] (1.47,0.5) -- (1.87,0.5);
    \draw[line width=1.0pt, dashed] (0.25,0.35) -- (1.55,0.35);
    \end{tikzpicture}\nonumber\\
&+&
\begin{tikzpicture}[anchor=base,baseline=8pt]
    \coordinate (A) at (0,0);
    \coordinate (B) at (2.,0);
    \coordinate (C) at (0,0.7);
    \coordinate (D) at (2.,0.7);
    \coordinate (D1) at (0,0);
    \coordinate (D2) at (1.,0);
    \coordinate (D3) at (1.3,0);
    \coordinate (D4) at (0.65,0.7);
    \coordinate (D5) at (1.6,0.7);
    \coordinate (D6) at (2.,0.7);
    \draw[line width=1.0pt] (A) -- (B);
    \draw[line width=1.0pt] (D) -- (C);
    \fill (D1) circle (2pt);
    \fill[white] (D1) circle (1pt);
    \fill (D2) circle (2pt);
    \fill (D3) circle (2pt);
    \fill (D4) circle (2pt);
    \fill (D5) circle (2pt);
    \fill[white] (D5) circle (1pt);
    \fill (D6) circle (2pt);
    \fill[white] (D6) circle (1pt);
    \draw[line width=1.0pt,snake it] (D1) -- (D4);
    \draw[line width=1.0pt,snake it] (D2) -- (D5);
    \draw[line width=1.0pt,snake it] (D3) -- (D6);
    \draw[line width=1.0pt, dashed] (0.25,0.35) -- (1.55,0.35);
    \end{tikzpicture}
    +  \!\ldots\!
    \nonumber
 \end{eqnarray}
    Again, the first contraction on the right-hand side reproduces the
    lowest-order diagram three times and the second contraction reproduces a combination of the first two diagrams in the previous diagram, and so on.

\subsection{Selfenergy}
Since all the diagrams are separable or inseparable~\cite{SchoellerSchoenKeldysh},
we can now define the selfenergy consisting of all inseparable diagrams (diagrams that cannot be cut into two by a vertical line without crossing a wiggled or a dashed line).
In the diagrammatic form it can be written as,
\begin{eqnarray}
  \Sigma &=&
    \begin{tikzpicture}[anchor=base,baseline=8pt]
    \coordinate (A) at (0,0);
    \coordinate (B) at (0.8,0);
    \coordinate (C) at (0,0.7);
    \coordinate (D) at (0.8,0.7);
    \draw[line width=1.0pt] (A) -- (B);
    \draw[line width=1.0pt,snake it] (A) -- (D);
    \draw[line width=1.0pt] (D) -- (C);
    \fill (A) circle (2pt);
    \fill (D) circle (2pt);
    \fill[white] (D) circle (1pt);
    \end{tikzpicture}\!+
  \!\ldots\!
  +\!
  \begin{tikzpicture}[anchor=base,baseline=8pt]
    \coordinate (A) at (0,0);
    \coordinate (B) at (1,0);
    \coordinate (C) at (0,0.7);
    \coordinate (D) at (1,0.7);
    \coordinate (E) at (0.6,0.0);
    \coordinate (F) at (0.3,0.7);
    \draw[line width=1.0pt] (A) -- (B);
    \draw[line width=1.0pt, snake it] (A) -- (D);
    \draw[line width=1.0pt, snake it] (E) -- (F);
    \draw[line width=1.0pt] (D) -- (C);
    \fill (A) circle (2pt);
    \fill[white] (A) circle (1pt);
    \fill (D) circle (2pt);    
    \fill (E) circle (2pt);
    \fill (F) circle (2pt);
    \fill[white] (F) circle (1pt);
    \end{tikzpicture}+\ldots\\
   & & +
   \begin{tikzpicture}[anchor=base,baseline=8pt]
    \coordinate (A) at (0,0);
    \coordinate (B) at (1.2,0);
    \coordinate (C) at (0,0.7);
    \coordinate (D) at (1.2,0.7);
    \coordinate (E) at (0.8,0.);
    \coordinate (F) at (0.3,0.0);
    \draw[line width=1.0pt] (A) -- (B);
    \draw[line width=1.0pt, snake it] (C) -- (F);
    \draw[line width=1.0pt, snake it] (E) -- (D);
    \draw[line width=1.0pt] (D) -- (C);
    \draw[line width=1.0pt, dashed] (0.17,0.35) -- (1,0.35);
    \fill (C) circle (2pt);
    \fill (D) circle (2pt);
    \fill (E) circle (2pt);
    \fill[white] (E) circle (1pt);
    \fill (F) circle (2pt);
    \fill[white] (F) circle (1pt);
    \end{tikzpicture}\!+\!\ldots\!
    +\!
   \begin{tikzpicture}[anchor=base,baseline=8pt]
    \coordinate (A) at (0,0);
    \coordinate (B) at (2.1,0);
    \coordinate (C) at (0,0.7);
    \coordinate (D) at (2.1,0.7);
    \coordinate (E) at (0.9,0.);
    \coordinate (F) at (0.3,0.0);
    \coordinate (G) at (1.2,0.7);
    \coordinate (H) at (1.5,0.7);
    \draw[line width=1.0pt] (A) -- (B);
    \draw[line width=1.0pt, snake it] (C) -- (F);
    \draw[line width=1.0pt, snake it] (E) -- (G);
    \draw[line width=1.0pt, snake it] (H) -- (B);
    \draw[line width=1.0pt] (D) -- (C);
    \draw[line width=1.0pt, dashed] (0.17,0.2) -- (1.05,0.2);
    \draw[line width=1.0pt, dashed] (1.1,0.41) -- (1.7,0.41);
    \fill (C) circle (2pt);
    \fill (G) circle (2pt);
    \fill (E) circle (2pt);
    \fill[white] (E) circle (1pt);
    \fill (F) circle (2pt);
    \fill[white] (F) circle (1pt);
    \fill (H) circle (2pt);
    \fill[white] (H) circle (1pt);
    \fill (B) circle (2pt);
    \end{tikzpicture}\ldots \nonumber
\end{eqnarray}
Using the selfenergy we can write the time evolution $\Pi(t_0,t)$ as 
\begin{eqnarray}\label{eq_rho_timeevoltuion_diagram}
  \Pi(t_0,t) &=&
   \begin{tikzpicture}[anchor=base,baseline=8pt]
    \coordinate (A) at (0,0);
    \coordinate (B) at (1,0);
    \coordinate (C) at (0,0.7);
    \coordinate (D) at (1,0.7);
    \draw[line width=1.0pt] (A) -- (B);
    \draw[line width=1.0pt] (D) -- (C);
    \end{tikzpicture}+
  \begin{tikzpicture}[anchor=base,baseline=8pt]
    \coordinate (A) at (0,0);
    \coordinate (B) at (1.2,0);
    \coordinate (C) at (0,0.7);
    \coordinate (D) at (1.2,0.7);
    \coordinate (E) at (0.3,0.0);
    \coordinate (F) at (0.3,0.7);
    \coordinate (G) at (0.9,0.0);
    \coordinate (H) at (0.9,0.7);
    \coordinate (I) at (0.6,0.3);
    \draw[line width=1.0pt] (A) -- (B);
    \draw[line width=1.0pt] (E) -- (F);
    \draw[line width=1.0pt] (G) -- (H);
    \draw[line width=1.0pt] (D) -- (C);
    \node  at (0.6,0.2) {$\Sigma$};
    \end{tikzpicture}
   +
  \begin{tikzpicture}[anchor=base,baseline=8pt]
    \coordinate (A) at (0,0);
    \coordinate (B) at (2,0);
    \coordinate (C) at (0,0.7);
    \coordinate (D) at (2,0.7);
    \coordinate (E) at (0.3,0.0);
    \coordinate (F) at (0.3,0.7);
    \coordinate (G) at (0.9,0.0);
    \coordinate (H) at (0.9,0.7);
    \coordinate (I) at (1.1,0.0);
    \coordinate (J) at (1.1,0.7);
    \coordinate (K) at (1.7,0.0);
    \coordinate (L) at (1.7,0.7);
    \coordinate (M) at (0.6,0.3);
    \draw[line width=1.0pt] (A) -- (B);
    \draw[line width=1.0pt] (E) -- (F);
    \draw[line width=1.0pt] (G) -- (H);
    \draw[line width=1.0pt] (D) -- (C);
     \draw[line width=1.0pt] (I) -- (J);
     \draw[line width=1.0pt] (K) -- (L);
    \node  at (0.6,0.2) {$\Sigma$};
     \node  at (1.4,0.2) {$\Sigma$};
    \end{tikzpicture} \\
    & & +
       \begin{tikzpicture}[anchor=base,baseline=8pt]
    \coordinate (A) at (0,0);
    \coordinate (B) at (2.8,0);
    \coordinate (C) at (0,0.7);
    \coordinate (D) at (2.8,0.7);
    \coordinate (E) at (0.3,0.0);
    \coordinate (F) at (0.3,0.7);
    \coordinate (G) at (0.9,0.0);
    \coordinate (H) at (0.9,0.7);
    \coordinate (I) at (1.1,0.0);
    \coordinate (J) at (1.1,0.7);
    \coordinate (K) at (1.7,0.0);
    \coordinate (L) at (1.7,0.7);
    \coordinate (M) at (1.9,0.0);
    \coordinate (N) at (1.9,0.7);
    \coordinate (O) at (2.5,0.0);
    \coordinate (P) at (2.5,0.7);
    \draw[line width=1.0pt] (A) -- (B);
    \draw[line width=1.0pt] (E) -- (F);
    \draw[line width=1.0pt] (G) -- (H);
    \draw[line width=1.0pt] (D) -- (C);
     \draw[line width=1.0pt] (I) -- (J);
     \draw[line width=1.0pt] (K) -- (L);
      \draw[line width=1.0pt] (M) -- (N);
     \draw[line width=1.0pt] (O) -- (P);
    \node  at (0.6,0.2) {$\Sigma$};
     \node  at (1.4,0.2) {$\Sigma$};
     \node  at (2.2,0.2) {$\Sigma$};
    \end{tikzpicture}+\ldots\nonumber\\
    &=&  \begin{tikzpicture}[anchor=base,baseline=8pt]
    \coordinate (A) at (0,0);
    \coordinate (B) at (1,0);
    \coordinate (C) at (0,0.7);
    \coordinate (D) at (1,0.7);
    \draw[line width=1.0pt] (A) -- (B);
    \draw[line width=1.0pt] (D) -- (C);
    \end{tikzpicture}+
  \begin{tikzpicture}[anchor=base,baseline=8pt]
    \coordinate (A) at (0,0);
    \coordinate (B) at (1.5,0);
    \coordinate (C) at (0,0.7);
    \coordinate (D) at (1.5,0.7);
    \coordinate (E) at (0.3,0.0);
    \coordinate (F) at (0.3,0.7);
    \coordinate (G) at (0.9,0.0);
    \coordinate (H) at (0.9,0.7);
    \coordinate (I) at (0.6,0.3);
    \draw[line width=1.0pt] (A) -- (B);
    \draw[line width=1.0pt] (E) -- (F);
    \draw[line width=1.0pt] (G) -- (H);
    \draw[line width=1.0pt] (D) -- (C);
    \draw[line width=1.0pt] (B) -- (D);
    \node  at (0.6,0.2) {$\Sigma$};
    \node  at (1.2,0.2) {$\Pi$};
    \end{tikzpicture}\nonumber
\end{eqnarray}
Using the above Dyson equation we get the equation of motion for the reduced density matrix 
\begin{equation}\label{eq:Dyson}
 \partial_t\hat{P}(t)=i[\hat{P}(t),H_{\rm S}]+\int_{t_0}^t\Sigma(t',t)\hat{P}(t')dt' ,
\end{equation}
which is in principle exact. Obviously, the key difficulty here is the calculation of the  selfenergy $\Sigma(t',t)$. 

\subsection{Leading-order master equation}
We will now derive the leading-order approximation for the equation of motion of the density matrix.
The selfenergy in the leading order is a sum of all eight second-order diagrams, which have the form
\begin{eqnarray}
  \Sigma &=&
    \begin{tikzpicture}[anchor=base,baseline=8pt]
    \coordinate (A) at (0,0);
    \coordinate (B) at (0.8,0);
    \coordinate (C) at (0,0.7);
    \coordinate (D) at (0.8,0.7);
    \draw[line width=1.0pt] (A) -- (B);
    \draw[line width=1.0pt,snake it] (A) -- (D);
    \draw[line width=1.0pt] (D) -- (C);
    \fill (A) circle (2pt);
    \fill (D) circle (2pt);
     \fill[white] (D) circle (1pt);
    \end{tikzpicture}
     +
    \begin{tikzpicture}[anchor=base,baseline=8pt]
    \coordinate (A) at (0,0);
    \coordinate (B) at (0.8,0);
    \coordinate (C) at (0,0.7);
    \coordinate (D) at (0.8,0.7);
    \draw[line width=1.0pt] (A) -- (B);
    \draw[line width=1.0pt,snake it] (B) -- (C);
    \draw[line width=1.0pt] (D) -- (C);
    \fill (B) circle (2pt);
    \fill (C) circle (2pt);
    \fill[white] (C) circle (1pt);
    \end{tikzpicture}
    +
     \begin{tikzpicture}[anchor=base,baseline=8pt]
    \coordinate (A) at (0,0);
    \coordinate (B) at (0.8,0);
    \coordinate (C) at (0,0.7);
    \coordinate (D) at (0.8,0.7);
    \draw[line width=1.0pt] (A) -- (B);
    \draw[line width=1.0pt,snake it]  (A) to [bend left=60] (B);
    \draw[line width=1.0pt] (D) -- (C);
    \fill (A) circle (2pt);
    \fill (B) circle (2pt);
    \fill[white] (B) circle (1pt);
    \end{tikzpicture}
    +  
    \begin{tikzpicture}[anchor=base,baseline=8pt]
    \coordinate (A) at (0,0);
    \coordinate (B) at (0.8,0);
    \coordinate (C) at (0,0.7);
    \coordinate (D) at (0.8,0.7);
    \draw[line width=1.0pt] (A) -- (B);
    \draw[line width=1.0pt,snake it] (C) to [bend right=60] (D);
    \draw[line width=1.0pt] (D) -- (C);
    \fill (C) circle (2pt);
    \fill[white] (C) circle (1pt);
    \fill (D) circle (2pt);
    \end{tikzpicture}\\
    & & + 
     \begin{tikzpicture}[anchor=base,baseline=8pt]
    \coordinate (A) at (0,0);
    \coordinate (B) at (0.8,0);
    \coordinate (C) at (0,0.7);
    \coordinate (D) at (0.8,0.7);
    \draw[line width=1.0pt] (A) -- (B);
    \draw[line width=1.0pt,snake it] (A) -- (D);
    \draw[line width=1.0pt] (D) -- (C);
    \fill (A) circle (2pt);
    \fill[white] (A) circle (1pt);
    \fill (D) circle (2pt);
    \end{tikzpicture}
     +
    \begin{tikzpicture}[anchor=base,baseline=8pt]
    \coordinate (A) at (0,0);
    \coordinate (B) at (0.8,0);
    \coordinate (C) at (0,0.7);
    \coordinate (D) at (0.8,0.7);
    \draw[line width=1.0pt] (A) -- (B);
    \draw[line width=1.0pt,snake it] (B) -- (C);
    \draw[line width=1.0pt] (D) -- (C);
    \fill (B) circle (2pt);
    \fill[white] (B) circle (1pt);
    \fill (C) circle (2pt);
    \end{tikzpicture}
    +
     \begin{tikzpicture}[anchor=base,baseline=8pt]
    \coordinate (A) at (0,0);
    \coordinate (B) at (0.8,0);
    \coordinate (C) at (0,0.7);
    \coordinate (D) at (0.8,0.7);
    \draw[line width=1.0pt] (A) -- (B);
    \draw[line width=1.0pt,snake it]  (A) to [bend left=60] (B);
    \draw[line width=1.0pt] (D) -- (C);
    \fill (A) circle (2pt);
    \fill[white] (A) circle (1pt);
    \fill (B) circle (2pt);
    \end{tikzpicture}
    +  
    \begin{tikzpicture}[anchor=base,baseline=8pt]
    \coordinate (A) at (0,0);
    \coordinate (B) at (0.8,0);
    \coordinate (C) at (0,0.7);
    \coordinate (D) at (0.8,0.7);
    \draw[line width=1.0pt] (A) -- (B);
    \draw[line width=1.0pt,snake it] (C) to [bend right=60] (D);
    \draw[line width=1.0pt] (D) -- (C);
    \fill (C) circle (2pt);
    \fill (D) circle (2pt);
     \fill[white] (D) circle (1pt);
    \end{tikzpicture}\nonumber
\end{eqnarray}
By evaluating each contribution explicitly we can write down the master equation.
As an example, we evaluate the contribution from the first diagram,
\begin{eqnarray}\label{eq:Diagram1}
&& \Sigma_{\bar s \bar s'\rightarrow ss'}(t',t) \nonumber \\
&&=  \langle \bar s' \vert \hat T \vert s'\rangle   \langle s \vert \hat T^\dagger \vert \bar s\rangle  P(t',t) e^{i(E_{s'}-E_{\bar s})(t-t')}. \nonumber
\end{eqnarray}
In the  Markov approximation, one neglects memory effects in the system time-evolution,
which means that in Eq.~(\ref{eq:Dyson}) the density matrix $\hat P(t')$ is replaced by the free evolution
$e^{iH_{\rm S}(t-t')}\hat P(t)e^{-iH_{\rm S}(t-t')}$.
This is a convenient but not a necessary approximation to be done here.
The Markov approaximation is equivalent to the analysis to of the crossing diagrams~\cite{Conditions_For_Markovianity}.
If the function $P(t_1,t_2)$ decays fast, such as in the high-temperature limit considered in Sec.~\ref{sec:CutoffSmallerThanTemperature},
the Markov approximation is well justified.
This allows for a straightforward integration of the equations over the time $t'$, giving
generalized transition rates of the form
\begin{eqnarray}
&&\Gamma(E)\equiv \int_{-\infty}^t dt' P(t',t) e^{-i E(t-t')}  \\
&&=\int_{-\infty}^t dt' e^{C(t'-t)} e^{i E(t'-t)} = \pi P(E) + iR(E).  \nonumber
\end{eqnarray}
Here, the first term on the right-hand side is purely real and corresponds to the $P(E)$-function
defined in Eq.~(\ref{eq:PEFunction}). This describes incoherent transitions with energy exchange with the
environment.
The second term is purely imaginary and corresponds to energy renormalization effects.
These are usually neglected or included into the system Hamiltonian.
They can result in important observable effects~\cite{Dima2016}.
This term can be written in the form
\begin{equation}
R(E)=-{\cal P}\int_{-\infty}^{\infty}d\omega\frac{P(E+\omega)}{\omega},
\end{equation}
where ${\cal P}$ indicates that the integration is made as a principal-value around $\omega=0$.

In total, after summing over all  the eight diagrams, one obtains
the well-known Bloch-Redfield equations of motion~\cite{Buch_Weiss,Shnirman_Review}
\begin{eqnarray}\label{eq:masterschoredinger}
   \dot P_{ss'}(t) & =& -i(E_s-E_{s'}) P_{s s'}(t) +\sum_{\bar s \bar s'}\Pi_{\bar s \bar s' s s}(t),
 \end{eqnarray}
where the performed unitary transformation affects to the form of the generalized transition rates
\begin{eqnarray}
  \Pi_{\bar s \bar s' s s}(t)           & = & \Gamma_{\bar s' s'}  \left [ T^\dagger_{s\bar s}  T_{\bar s' s'} + T_{s\bar s}  T^\dagger_{\bar s' s'} \right] P_{\bar s\bar s'}(t)  \nonumber \\
             & + & \Gamma_{\bar s s}^*  \left [ T^\dagger_{\bar s s}  T_{\bar s' s'} + T_{\bar s s}  T^\dagger_{\bar s' s'} \right] P_{\bar s\bar s'}(t)  \nonumber \\
              & - & \sum_v\Gamma_{\bar sv}^*  \left [T^\dagger_{sv} T_{v \bar s} +T_{sv} T^\dagger_{v\bar s}\right] P_{\bar s\bar s'}(t) \delta_{\bar s' s'}  \nonumber \\
              & - &\sum_v \Gamma_{\bar s'v} \left [ T^\dagger_{\bar s' v} T_{v s'} + T_{\bar s' v} T^\dagger_{v s'}\right]   P_{\bar s\bar s'}(t) \delta_{\bar s s} \nonumber ,
\end{eqnarray}
with $T_{ij}\equiv \langle i \vert \hat T \vert j\rangle $ and $\Gamma_{ij}\equiv \Gamma(E_i)-\Gamma(E_j)$.

The leading-order master equation under the polaron transformation is a useful and straightforward tool for many systems. Our previous studies include
lasing under strong noise~\cite{JC_1,JC_3} and incoherent Cooper-pair tunneling in Josephson junction arrays~\cite{JaredJJarray}.
The diagrammatic formulation of higher orders derived in the preceding sub-sections then allows us to study convergence criteria,
performed in Sec.~\ref{sec:TheNoiseSpectra}.

\section{Strong low-frequency noise and conditions for convergence}\label{sec:TheNoiseSpectra}

 In the preceding Section, we presented a diagrammatic expansion
 of the time evolution of the density matrix and derived the master equation in the leading order.
 Here, we want to discuss the conditions for the leading-order expansion to be valid. For this we first want to introduce a specific 
 model for the spectral function. 

We consider a case which is often discussed within $P(E)$-theory~\cite{pofe_Nazarov},
where the spectral function can be related to an impedance $Z(\omega)$ via
\begin{equation}
 J(\omega)=2\pi\omega \frac{{\rm Re} [Z(\omega)]}{R_{\rm Q}}
\end{equation}
where $R_{\rm Q}=h/e^2$ is the resistance quantum.
If our system has a dipole moment and is capacitively coupled to 
an Ohmic environment, the impedance can take the form
\begin{equation}
 {\rm Re} [Z(\omega)]=\frac{R}{1+(\omega RC)^2}
\end{equation}
Here $R$ characterizes the dissipation of the environment and the capacitance $C$ defines a cut-off frequency $\omega_R=1/RC$.
The spectral density in this case can also be written as,
\begin{equation}\label{eq_Ohmic_spectral_density}
 J(\omega)=2\epsilon_C \omega \frac{\omega_R}{\omega^2+\omega_R^2}
\end{equation}
with the charging energy $\epsilon_C=e^2/2C$.


\subsection{Cut-off freqencies smaller than temperature}\label{sec:CutoffSmallerThanTemperature}
If we consider the occupation of the modes by finite temperature $k_B T$, for small cut-off frequencies, $\omega_R\ll k_B T$,
we get a spectral function,
\begin{equation}\label{eq_lorentz_spectral_function}
 S(\omega)=J(\omega)\coth(\omega/2 k_B T)\approx \frac{4\epsilon_C k_B T \omega_R}{\omega^2+\omega_R^2}\,.
\end{equation}
This spectral function has a maximum at $\omega=0$ and therefore the noise we are considering is low frequency noise.
The spectral function is characterized by height $\epsilon_C k_{\rm B}T/\omega_R$ 
and width $\omega_R$.
This can be an important regime, even at milli-Kelvin temperatures \cite{dwave_noise_measurment,Leppakangas2015,JaredJJarray}, since cut-off frequencies can be even smaller.

\subsubsection{Contraction function}\label{Contraction_function}
We study now further the spectral density of Eq.~(\ref{eq_Ohmic_spectral_density}) in the limit $\omega_R \ll k_B T$.
The corresponding contraction function $P(t_1,t_2)$, see Eqs.~(\ref{eq:Correlator}-\ref{eq:Correlator2}), has the form
\begin{eqnarray}
&& P(t_1,t_2)= \label{eq:ContractionHighTemperature} \\
&&\exp\left\{ -\frac{2c^2\epsilon_C k_B T\left(1-e^{-\omega_R \vert \delta t\vert}-\omega_R\vert \delta t\vert\right)}{\omega_R^2}\right.\nonumber \\
& & \left.-i\frac{2c^2\epsilon_C(1-e^{-\omega_R \vert \delta t\vert })}{\omega_R}{\rm Sign}(\delta t)\nonumber\,.
\right\}
\end{eqnarray}
Here we have defined $\delta t=t_1-t_2$.
The real part behaves at short times quadratically and at long times linearly.
A characteristic time for the cross over is $1/\omega_R=RC$.
The result for the imaginary part does not depend on temperature and is general.

We will now use the coupling $c=1$, for other values all the results can be obtained by the change $\epsilon_C\rightarrow c^2\epsilon_C$.
In the strong-coupling limit, $\omega_R \ll \sqrt{\epsilon_C k_B T}$, we can use the short time approximation and the relevant behaviour simplifies to\cite{Stron_resistance_Grabert}
\begin{equation}\label{eq_Simplified_contraction}
 P(t_1,t_2)\approx \exp\left\{-\epsilon_C \left[ k_B T (t_1-t_2)^2+i(t_1-t_2)\right]\right\}.
\end{equation}
This also implies that we consider a spectral function which is sharply peaked at small frequencies. 

We note that when using only quadratic approximation for correlation fucntions, the corresponding coupling between the pairs ($F$-function) does not go to zero with increasing the distance
     between pairs. However, it does go to zero when the linear behaviour of Eq.~(\ref{eq:ContractionHighTemperature}) dominates ($t>RC$). This is
     also seen in the convergence results derived below.

It should also be noted that there is a direct relation between the spectral function (\ref{eq_lorentz_spectral_function})
and $1/f$ noise \cite{Shnirman_Two_to_one_over_f}.
It is known that $1/f$ noise can be described by many superimposed Lorentzian spectra, with a probability distribution for the width $\omega_R$
which is given by $1/\omega_R$,
\begin{equation}
 S_{1/f}(\omega)=\int_0^{\infty}\!\!\! d\omega_R \frac{1}{\omega_R} S(\omega)=\frac{2\pi \epsilon_C k_B T}{\omega}
\end{equation}
As we will see later, the expansion we discuss here can accommodate low frequency noise of the type described by
Eq.~(\ref{eq_lorentz_spectral_function}) for rather small cut-off frequencies $\omega_R$, but not for $\omega_R\rightarrow 0$.

\subsubsection{Analysis of higher-order diagrams}\label{sec:HigherOrderDiagrams}
We assume strong coupling to the environment and therefore we assume
large $\sqrt{k_{\rm B}T\epsilon_C}$ compared to $\omega_R$.
In this case, $ P(t_1,t_2)$ decays quickly and allows for the Markov approximation in the leading-order master equation. However, in contrast to the expansion studied in e.g.~Ref.~\onlinecite{Conditions_For_Markovianity},
we have different classes of higher order diagrams which decay in different ways.

 Here, we analyze higher-order diagrams to understand the convergence conditions of our expansion. 
 At first we will consider a standard diagram with two crossed contractions,
 \begin{eqnarray}\label{eq_Crossing_Contractions_Inital}
  \begin{tikzpicture}[anchor=base,baseline=8pt]
  \node (N1) at (0.7,-0.4) {$t_1$};
  \node (N2) at (0.,-0.4) {$t'$};
  \node (N3) at (0.3,0.85) {$t_2$};
  \node (N4) at (1,0.85) {$t$};
    \coordinate (A) at (0,0);
    \coordinate (B) at (1,0);
    \coordinate (C) at (0,0.7);
    \coordinate (D) at (1,0.7);
    \coordinate (E) at (0.6,0.0);
    \coordinate (F) at (0.3,0.7);
    \draw[line width=1.0pt] (A) -- (B);
    \draw[line width=1.0pt, snake it] (A) -- (D);
    \draw[line width=1.0pt, snake it] (E) -- (F);
    \draw[line width=1.0pt] (D) -- (C);
    \fill (A) circle (2pt);
    \fill[white] (A) circle (1pt);
    \fill (D) circle (2pt);    
    \fill (E) circle (2pt);
    \fill (F) circle (2pt);
    \fill[white] (F) circle (1pt);
    \end{tikzpicture}=m^4\int_{t'}^t dt_1\int_{t'}^{t_1} dt_2 P(t,t_2)P(t',t_1) 
    \end{eqnarray}
    Here, the term $P(t,t_2)$ corresponds to the upper pair and $P(t',t_1)$ to the lower one.
    Here we made several assumptions. We assume that the relevant system energy splittings are small.
    If a system energy splitting is large, we get additional oscillating functions
    which improve convergence, therefore we are considering the worst-case scenario. The prefactor is given by 
    the assumption that there is a characteristic energy scale $m$ which corresponds to the relevant matrix element 
    $m\propto \langle n|\hat{T}|n'\rangle$ where $|n\rangle$ and $|n'\rangle$
     are eigenstates of the system.

     In a rough approximation, where we also neglect the oscillating part in the contraction
     of Eq.~(\ref{eq_Simplified_contraction}) and assuming that $t-t'$ is relatively large, we find 
    \begin{eqnarray}\label{eq_Crossing_Contractions_Final}
  \begin{tikzpicture}[anchor=base,baseline=8pt]
    \coordinate (A) at (0,0);
    \coordinate (B) at (1,0);
    \coordinate (C) at (0,0.7);
    \coordinate (D) at (1,0.7);
    \coordinate (E) at (0.6,0.0);
    \coordinate (F) at (0.3,0.7);
    \draw[line width=1.0pt] (A) -- (B);
    \draw[line width=1.0pt, snake it] (A) -- (D);
    \draw[line width=1.0pt, snake it] (E) -- (F);
    \draw[line width=1.0pt] (D) -- (C);
    \fill (A) circle (2pt);
    \fill[white] (A) circle (1pt);
    \fill (D) circle (2pt);    
    \fill (E) circle (2pt);
    \fill (F) circle (2pt);
    \fill[white] (F) circle (1pt);
    \end{tikzpicture}\approx 
    -\sqrt{\frac{\pi}{2}}
    \frac{m^4 e^{-\frac{1}{2}\epsilon_C k_{\rm B} T (t-t')^2}}{(\epsilon_C k_{\rm B} T)^{3/2}(t-t')}
    \end{eqnarray}
         If we now compare the size of the lowest-order diagram with this results we see that we need,
        \begin{equation}
         \frac{m^2}{\sqrt{\epsilon_C k_B T}}\gg \frac{m^4}{(\epsilon_C k_B T)^{3/2}}\,.
        \end{equation}
  This gives us the rule for convergence,
  \begin{equation}
   \frac{m}{\sqrt{\epsilon_C k_B T}}\ll 1.
  \end{equation}
  Basically we see that the higher-order diagram of the form~(\ref{eq_Crossing_Contractions_Final})
  can be neglected if the contraction $P(t_1,t_2)$ has a decay rate 
  which is much larger than the coupling  constant $m$.
  This is a well known rule which applies for many expansion theories.

  In the limit we are considering, at first sight
  it seems as if the cut-off frequency of the spectral function plays no role for the convergence.  
  However, this is essential for different set of diagrams, which have no crossing contractions
  but are inseparable because of a connector. These terms have the form
    \begin{eqnarray}\label{eq_4th_order_diag_with_connector}
    \begin{tikzpicture}[anchor=base,baseline=8pt]
     \node (N1) at (0.3,-0.4) {$t_2$};
     \node (N2) at (1.2,-0.4) {$t$};
     \node (N3) at (0,0.85) {$t'$};
     \node (N4) at (0.8,0.85) {$t_1$};
    \coordinate (A) at (0,0);
    \coordinate (B) at (1.2,0);
    \coordinate (C) at (0,0.7);
    \coordinate (D) at (1.2,0.7);
    \coordinate (E) at (0.8,0.7);
    \coordinate (F) at (0.3,0.0);
    \draw[line width=1.0pt] (A) -- (B);
    \draw[line width=1.0pt, snake it] (C) -- (F);
    \draw[line width=1.0pt, snake it] (E) -- (B);
    \draw[line width=1.0pt] (D) -- (C);
    \draw[line width=1.0pt, dashed] (0.17,0.35) -- (1,0.35);
    \fill (C) circle (2pt);
    \fill (E) circle (2pt);
    \fill (B) circle (2pt);
    \fill[white] (B) circle (1pt);
    \fill (F) circle (2pt);
    \fill[white] (F) circle (1pt);
    \end{tikzpicture}
    \!\!\!=\! m^4\!\int_{t'}^t\!\! dt_1\int_{t'}^{t_1}\!\!\!\! dt_2
    P(t,t_1)P(t_2,t')F(t_2,t,t_1,t') \nonumber \\
    \end{eqnarray}
    Diagrams of this form have also been shown to be relevant for the calculation of the statistics of photon emission in 
    voltage biased Josephson junction \cite{Leppakangas2015,Leppakangas2016}.
	Here, the function $P(t,t_1)$ describes the timewise later pair and the function $P(t_2,t')$
	the earlier one, whereas $F(t_2,t,t_1,t')$ the interaction between the two.
    We know that the functions $P(t_1,t)$ and $P(t',t_2)$ decay very fast. Therefore, we 
    expand the connector $F$ around $t_1=t$ and $t_2=t'$ using the Taylor expansion.
    This is again a valid approximation for relatively large $t-t'$.
The lowest-order nonzero element of this expansion is given by,
\begin{eqnarray}
 &&\left.\frac{\partial F}{\partial t_1\partial t_2}\right|_{t_1=t,t_2=t'} = -\frac{1}{\pi}\int_0^{\infty} \nonumber
 d\omega J(\omega) \coth\left(\frac{\omega}{2k_B T} \right) \\
 &&\times \left[ \cos \omega (t-t')  -i \sin \omega (t-t') \right].
\end{eqnarray}
In this approximation we can write the diagram~(\ref{eq_4th_order_diag_with_connector})
in the form
  \begin{eqnarray}
    \begin{tikzpicture}[anchor=base,baseline=8pt]
    \coordinate (A) at (0,0);
    \coordinate (B) at (1.2,0);
    \coordinate (C) at (0,0.7);
    \coordinate (D) at (1.2,0.7);
    \coordinate (E) at (0.8,0.7);
    \coordinate (F) at (0.3,0.0);
    \draw[line width=1.0pt] (A) -- (B);
    \draw[line width=1.0pt, snake it] (C) -- (F);
    \draw[line width=1.0pt, snake it] (E) -- (B);
    \draw[line width=1.0pt] (D) -- (C);
    \draw[line width=1.0pt, dashed] (0.17,0.35) -- (1,0.35);
    \fill (C) circle (2pt);
    \fill (E) circle (2pt);
    \fill (B) circle (2pt);
    \fill[white] (B) circle (1pt);
    \fill (F) circle (2pt);
    \fill[white] (F) circle (1pt);
    \end{tikzpicture}
    \!\!\! &=& \! -m^4\left.\frac{\partial^2 F}{\partial t_1\partial t_2}\right|_{t_1=t,t_2=t'} \times
    \\
        & &
     \!\int_{t'}^t\!\! dt_1\int_{t'}^{t_1}\!\!\!\! dt_2
     P(t,t_1)(t-t_1) P(t_2,t')(t_2-t'). \nonumber
    \end{eqnarray}
Integration over the times gives us
\begin{eqnarray}
&& \lim_{t-t' \to \infty}\int_{t'}^t\!\! dt_1\int_{t'}^{t_1}\!\!\!\! dt_2 P(t_1,t)(t_1-t) P(t',t_2)(t_2-t') \nonumber\\
&&= \frac{1}{ \epsilon_C^2 (k_{\rm B} T)^2}  .  
\end{eqnarray}
Using the approximation for our spectral density in Eq.~(\ref{eq_Ohmic_spectral_density})
as specified in Eq.~(\ref{eq_lorentz_spectral_function}),
we can analytically estimate the contribution from the connector,
\begin{equation}\label{eq_connecter_decay}
\left. \frac{\partial^2 F}{\partial t_1\partial t_2}\right|_{t_1=t,t_2=t'}
 =2\epsilon_C k_{\rm B} T e^{-\omega_R (t-t')}.
\end{equation}
This analysis implies that for relatively large times $t-t'$ the connector decays as the memory ($RC$) time of the environment.
From this we find the order of magnitude of the contribution of the diagram~(\ref{eq_4th_order_diag_with_connector}),
which becomes $m^4/\epsilon_C k_{\rm B} T \omega_R$.
If we compare this to the contribution of the lowest order diagram, we find the convergence rule
\begin{equation}\label{eq:ConvergenceCondition2}
 \frac{m^2}{\omega_R\sqrt{\epsilon_C k_{\rm B} T}}\ll 1.
\end{equation}
From Eq.~(\ref{eq:ConvergenceCondition2}) we see that a noise source with a rather small cut-off frequency can in fact be treated,
as long as the overall noise magnitude is sufficiently large.
We also see that the in limit $\omega_R\rightarrow 0$ the expansion does not converge.
This is natural since in this limit the memory time of the environment approaches infinity.
In summary, we see that  also the bath cut-off frequency plays an essential role in
convergence of the expansion. 

\subsection{Convergence for temperatures smaller than cut-off frequency}

Within similar analysis we can also study the convergence in the case of strong coupling of the environment and small cut-off frequencies 
 but temperatures even smaller then the cut-off frequency, $\omega_R>k_B T$. The width of the spectral function is characterized now by $k_{\rm B}T$,
meaning that strong coupling limit corresponds to $\epsilon_C \gg \omega_R$, which is equivalent to $R\gg R_{\rm Q}$.
In this case the result for the contraction function $P(t_1,t_2)$ can be derived from the formally exact solution given 
in Ref.~\onlinecite{exact_solution}. We find a new solution for the real part in the short and long time limits,
\begin{widetext}
\begin{equation}
 {\rm Re}[P(t_1,t_2)]=\left\{ \begin{array}{cc}
                     \exp\left[-\frac{\epsilon_C}{\pi\omega_R} \left( \gamma - \cosh(\omega_R t){\rm chi}(\omega_R t) + \sinh(\omega_R t){\rm shi}(\omega_R t) \right) \right] & k_B T t\ll 1\\
                     \exp\left[\frac{\epsilon_C k_B T}{\omega_R} t 
                     -\frac{\epsilon_C\left(H_{-\frac{\omega_R}{2 k_B T\pi}}
                       +H_{\frac{\omega_R}{2 k_B T\pi}}-\pi\cot(\omega_R/2 k_B T) \right)}
                       {2\pi\omega_R} \right] & k_B T t \gg 1                    
                    \end{array}\right.
\end{equation}
\end{widetext}
where $H_n$ is the Harmonic Number and ${\rm chi}(x)$ and ${\rm shi}(x)$ are the $\cosh$ and $\sinh$ integrals respectively.
For $T=0$, the short time limit is in fact the correct result at all times. In this case our expansion will diverge and other methods \cite{Kashuba_all_orders,Saptsov_all_orders}
need to be used. However, for any finite temperature the long time limit holds. 
For the considered limit $\omega_R>k_B T$, Eq.~(\ref{eq_connecter_decay}) stays the same, and all diagrams decay in the same way. The convergence analysis gives in this case only one condition
\begin{equation}
 \frac{m \omega_R}{\epsilon_C k_B T}\ll 1.
\end{equation}
We then obtain that even at very small temperatures, convergence can be achived in the strong coupling limit $\epsilon_C \gg \omega_R$.

\section{Conclusions}\label{sec:Conclusions}

We discussed a master-equation expansion where first the coupling to the bath is 
diagonalized explicitly and then expanded in the system operators dressed by the bath operators. 
The motivation here is to study expansion schemes that can be used in the case of strong coupling to the environment.
We formally introduced contraction rules which allow for the division of all resulting terms in the expansion
into two time correlators of type $P(t_1,t_2)$ and four time correlators, described by the connector $F(t_1,t_2,t_3,t_4)$.
The introduced rules allowed for a consistent definition of the selfenergy.
We showed that the contribution of the connector
is important since it contains the effects of the slowly decaying correlations.
We then derived explicit limits when the leading-order master-equation approach is valid,
i.e.~when the contribution from the two types of diagrams in higher orders stays small.
The results clarify the limits of our previous works on lasing in systems under strong noise~\cite{JC_1,JC_3}, incoherent Cooper-pair tunneling in Josephson junction arrays~\cite{JaredJJarray}, as well as 
inelastic Cooper-pair tunneling across voltage-biased Josephson junctions
in the Coulomb blockade regime~\cite{Leppakangas2015,Leppakangas2016}.

As we discussed in section \ref{Contraction_function}, interesting is
the connection of the used noise spectral function to $1/f$-noise. Given a noise spectral function 
of Eq.~(\ref{eq_lorentz_spectral_function}), we can describe the high-frequency tail of $1/f$-noise,
\begin{eqnarray}
 S_{1/f,{\rm high}}&=&\int_{\omega_{R,{\rm min}}}^{\infty}d\omega_R \frac{1}{\omega_R}S(\omega) \\
 &=&\frac{2\epsilon_C k_B T\left[\pi-2 \arctan(\omega_{R,{\rm min}}/\omega)\right]}{\omega} \nonumber
\end{eqnarray}
where the low frequency limit for the cut-off frequency $\omega_R$ is determined by
the condition $m^2/\sqrt{\epsilon_C k_{\rm B} T} \omega_{R,{\rm min}}\ll 1$.
Such connection can then be used to theoretically account for a large part of low-frequency noise, in limits where traditional direct system-bath coupling expansions do not work.

\end{document}